\documentstyle[psfig,amssym]{elsart}

\begin{document}

\begin{frontmatter}

\title{TeV gamma rays from PSR 1706-44}

\author{P. M. Chadwick}, \author{M. R. Dickinson}, \author{N. A.
Dipper}, \author{J. Holder}, \author{T. R. Kendall}, \author{T. J. L.
McComb}, \author{K. J. Orford}, \author{J. L. Osborne}, \author{S. M.
Rayner}, \author{I. D. Roberts}, \author{S. E. Shaw} and \author{K. E.
Turver}

\address{Department of Physics, Rochester Building, Science
Laboratories, University of Durham, South Road, Durham DH1 3LE, UK}

\date{ and accepted for publication in {\it Astroparticle Physics}}

\begin{abstract}

Observations made with the University of Durham Mark 6 atmospheric
\v{C}erenkov telescope confirm that PSR B1706-44 is a very high energy
$\gamma$-ray emitter. There is no indication from our dataset that the
very high energy $\gamma$-rays are pulsed, in contrast to the findings
at $< 20$ GeV, which indicate that more than 80\% of the flux is pulsed.
The flux at $E > 300$ GeV is estimated to be $(3.9 \pm 0.7_{\rm stat})
\times 10^{-11}~{\rm cm}^{-2}~{\rm s}^{-1}$.

\end{abstract}

\begin{keyword}
VHE $\gamma$-ray astronomy -- pulsars -- PSR B1706-44

PACS codes: 95.85.Pw; 97.60.Gb; 98.38.Mz; 98.70.Rz
\end{keyword}

\end{frontmatter}

\section{Introduction} 

\subsection{PSR B1706-44} 

PSR B1706-44 is a 102 ms radio pulsar which may be associated with the
supernova remnant G343.1-2.3. The pulsar was discovered via its radio
emission \cite{John92}. Observations of the 102 ms periodicity with the
EGRET detector on the {\em Compton Gamma Ray Observatory\/}
\cite{Thom92,Thom96} associated the {\em COS-B\/} $\gamma$-ray source
2CG 342-02 \cite{Swan81} with this pulsar; it is one of only seven
identified pulsars which emit high energy $\gamma$-rays. Becker et al.
\cite{Beck94} showed that PSR B1706-44 also emits X-rays, which may be
an indication of the presence of a synchrotron X-ray nebula. McAdam et
al. \cite{Maca93} have suggested that PSR B1706-44 is associated with
the shell-type supernova remnant G343.1-2.3 (but see \cite{Nica96} for
an opposing view). Recent observations of candidate optical counterparts
in the PSR B1706-44 field have yielded no evidence for pulsations,
suggesting that the optical counterpart has an R-band magnitude of $< 18$
\cite{Chak98}.

A number of models of $\gamma$-ray emission from pulsars suggest that
PSR B1706-44, as well as the Crab, should be a source of very high
energy (VHE) $\gamma$-rays as there are a number of similarities between
the two objects (see, for example, \cite{Chen94}). The Crab nebula, with
its associated pulsar, is a well established source of VHE
$\gamma$-rays, detectable from a few hundred GeV \cite{Week89,Vaca91} to
higher energies \cite{Bail93,Tani94,Kono96,Tani97}. The Crab pulsar has
also been detected in VHE $\gamma$-rays \cite{Gibs82,Dowt84,Acha92},
though later observations using imaging telescopes with greater
sensitivity, and operating at lower threshold energy, have failed to
show evidence for pulsed emission \cite{Gill97}.

Kifune et al. \cite{Kifu95}, using the CANGAROO telescope, found
evidence of TeV $\gamma$-ray emission from PSR B1706-44 but detected no
pulsations. We have observed this object with the University of Durham
Mark 6 $\gamma$-ray telescope to extend measurements closer to the
energy range of the EGRET measurements.
 
\subsection{The University of Durham Mark 6 VHE $\gamma$-ray telescope}

The University of Durham Mark 6 VHE $\gamma$-ray telescope is situated
at Narrabri, NSW, Australia, 200m above sea level. It is an atmospheric
\v{C}erenkov telescope; such telescopes detect optical \v{C}erenkov
photons emitted as air showers produced by VHE $\gamma$-rays and cosmic
rays pass through the upper atmosphere. The Mark 6 telescope uses a
triple mirror, fast coincidence system, each mirror having an area of
$42~{\rm m}^{2}$ and a focal length of 7 m. There is a conventional
imaging camera consisting of 91~1 inch and 18~2 inch PMTs at the focus
of the centre mirror. At the focus of each of the left and right mirrors
is a low-resolution triggering camera consisting of 19 hexagonal PMTs.
Each 1 inch PMT in the imaging camera has a field of view of
$0.25^{\circ}$, giving a total field of view of 7 square degrees. The
telescope is described in detail by Armstrong et al. \cite{Arms97}. It
is triggered by a 3-fold spatial plus 4-fold temporal coincidence
\cite{Chad97}, which consist of two elements:

\begin{enumerate}

\item The first requirement is that a signal is detected from each of
a pair of PMTs viewing the same area of sky in the left and right detectors.

\item The second requirement is that any 2 of the 7 camera PMTs which
cover the same area of sky as the left/right PMTs which have responded
also produce a signal and that the 2 PMTs are adjacent.

\end{enumerate} 

This novel coincidence system has a number of advantages over a single
dish telescope, and provides both a very effective suppression of noise
due to the night sky background and a complete elimination of events due
to local muon triggers. It also gives the lowest practical energy
threshold for a given mirror area. We estimate that the telescope, in
its present configuration, can reliably detect $\gamma$-rays with
energies as low as 125 GeV near the zenith, although the detection
efficiency at this energy is $\leq 10\%$ of the peak detection
efficiency . An initial simulation study performed for the telescope
inclined at an angle of $20^{\circ}$ suggests that the peak differential
detection rate for $\gamma$-rays occurs at 250 GeV, assuming a source
with a power law spectral index of $-2.4$. Following recent convention, this is
taken as the energy threshold of the telescope, but it should be noted
that this is the result of a preliminary study only and the systematic
error at large zenith angles is estimated to be 50\%. For the
observations of PSR B1706-44, with a typical zenith angle of
$\sim~30^{\circ}$, the $\gamma$-ray energy threshold is estimated to be
300 GeV.

The telescope's detectors are calibrated during observations by use of a
nitrogen laser coupled to a plastic scintillator, which provides a
diffuse source of blue light at the centre of each mirror, thus
producing a pool of blue light which is uniform over the face of the
detector packages. The laser pulses are separated by random intervals
with a mean rate of 50 ${\rm min}^{-1}$ throughout all observations. The
pedestals of the charge digitisers associated with each PMT channel are
also measured throughout observations by means of randomly occuring
`null' events, also with a mean rate of 50 ${\rm min}^{-1}$. In
addition, the telescope is equipped with a comprehensive performance
monitoring system, whereby the PMTs' anode currents and noise rates are
recorded throughout observations, and environmental conditions are
monitored. Event arrival times are measured using a rubidium
oscillator-based clock, monitored against a GPS time signal, to a
relative accuracy of $1~\mu{\rm s}$ and an absolute epoch error of
$10~\mu{\rm s}$.

\section{Observations}

Observations of PSR B1706-44 were made on clear moonless nights at
zenith angles between $15^{\circ}$ and $50^{\circ}$ in May and July
1996. More than 90\% of the data were taken between culmination (a
zenith angle of $14^{\circ}$) and $35^{\circ}$. An observing log is
shown in Table~\ref{observing_log}. Our method of observation is to make
15 minute exposures on and off source, with an equal total exposure time
for each. The total exposure time on source, allowing for dead-time
during chopping manoeuvres, was 560 minutes.

\begin{table}

\caption{Log for observations of PSR B1706-44 with the
University of Durham Mark 6 telescope at Narrabri during
1996.}\label{observing_log}

\vspace{0.1cm}
\begin{center}
\begin{tabular}{@{}lrlr}
\hline\hline
Date&No. of&Date&No. of\\ 
(1996)&scans&(1996)&scans\\
\hline
May 11&3&July 9&7\\
May 12&4&July 11&4\\
May 21&5&July 15&4\\
May 22&2&July 18&5\\
May 24&6&&\\
\hline
\hline
\end{tabular}
\end{center}
\end{table}

Data were selected for analysis if pairs of ON and OFF observations have
raw count rates which differ by $\leq 2.5~\sigma$ and extensive
environmental monitoring showed clear and stable skies. The 40 scans
which passed this quality threshold are listed in Table~\ref{observing_log}.

\section{Analysis procedure}

The `raw' event totals shown in Table~\ref{results} constitute
\emph{all} the events recorded by the Mark 6 during the observations of
PSR B1706-44 and therefore include events near the telescope trigger
threshold and events of all $SIZE$s which fall near the edge of the
camera, as well as events more suitable for conventional imaging
analysis. Some of the out-of-geometry events recorded during the ON
source observations are likely to be due to the presence of a star
($\eta$ Scorpii) near the edge of the field of view. Events considered
suitable for analysis are those events which are confined within the
sensitive area of the camera and which contain sufficient information
for reliable image analysis, i.e. which have {\em SIZE\/} of 200 --
20000 digital counts, where 3 digital counts $ \sim 1$ photoelectron,
and 200 digital counts are produced by a 125 GeV $\gamma$-ray at the
zenith. This selection has the effect of removing any out-of-geometry
events induced by $\eta$ Scorpii together with other events which are
considered to contain insufficient information for analysis. The total
number of events after this process is included in the `$SIZE/DISTANCE$
selected' row of Table~\ref{results}.

Monte Carlo simulations indicate that the image shape of the
\v{C}erenkov light from a $\gamma$-ray shower can be approximated by an
ellipse, the major axis of which is oriented towards the source
position, whereas a cosmic ray (hadronic) shower produces a broader,
more irregularly shaped image \cite{Hill85}. The image can be
parameterised using techniques developed by the Whipple group which
describe both the shape and the orientation of the image. In addition, a
measure of the fluctuations between the samples recorded by the left and
right flux collectors of the Mark 6 telescope provides a further
discriminant \cite{Chad97}. $\gamma$-rays are identified on the basis of
image shape and left/right fluctuation, and then plotting the number of
events as a function of the pointing parameter {\em ALPHA}; $\gamma$-ray
events from a point source will appear as an excess of events at small
values of {\em ALPHA}. The cuts applied to the data are shown in
Table~\ref{cuts}.

\begin{figure} 
\centerline{\psfig{file=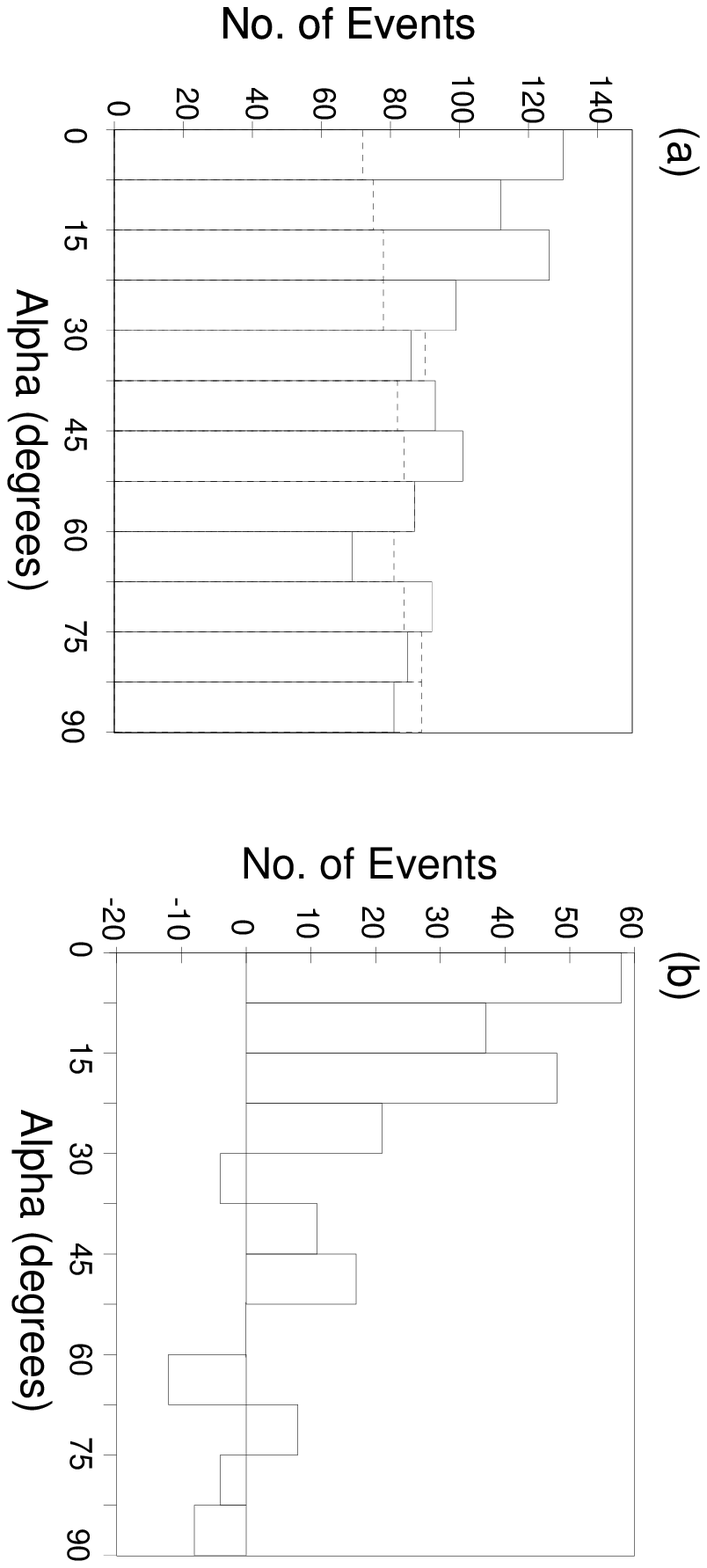,height=6cm,angle=90}}

\caption{ (a). The {\em ALPHA\/} distributions ON and OFF source for PSR
B1706-44. The dotted line refers to OFF source data. (b). The difference
between ON and OFF source data.}\label{alpha}

\end{figure}

The number of events remaining ON and OFF source after the application
of the cuts described above is summarized in Table~\ref{results}. The
{\em ALPHA\/} distributions of the ON and OFF source events have been
plotted, and the results are shown in Figure~\ref{alpha}. No
normalization of the ON and OFF data has been applied. There is an
excess of events at small {\em ALPHA\/}, the expected $\gamma$-ray
domain, and imposing a $\gamma$-ray cut of ${\em ALPHA\/} <
22.5^{\circ}$ yields a $\gamma$-ray detection significant at the
$5.9~\sigma$ level. There may be scope to decrease the width of the
alpha plot peak by correcting our data for the effects of such factors
as the Earth's geomagnetic field and the optical performance of the
telescope mirrors \cite{Chad98}.

\begin{table}

\caption{The results of selecting events on the basis of their shape and
value of {\em ALPHA}.}\label{results}

\vspace{0.1cm}
\begin{center}
\begin{tabular}{@{}lrrrr}
\hline\hline
&ON&OFF&Difference&Significance\\
\hline
Raw&251412&248405&3007&$4.3~\sigma$\\
$SIZE/DISTANCE$ Selected&130295&129380&915&$1.8~\sigma$\\
Shape Selected&1161&989&172&$3.7~\sigma$\\
Shape and {\em ALPHA\/} $< 22.5^{\circ}$&368&225&143&$5.9~\sigma$\\
\hline
\hline
\end{tabular}
\end{center}
\end{table}

\begin{table}

\caption{The cuts applied to the data from PSR B1706-44.}\label{cuts}

\vspace{0.1cm}
\begin{center}
\begin{tabular}{@{}lcc}
\hline\hline
Parameter&Limits (high energy events)&Limits (low energy events)\\
\hline
$SIZE$&$800-20000$ d.c.&$200-800$ d.c.\\
$DISTANCE$&$0.35^{\circ}-0.75^{\circ}$&$0.35^{\circ}-0.75^{\circ}$\\
$ECCENTRICITY$&$0.35-0.75$&---\\
$WIDTH$&$ < 0.26^{\circ}$&$ < 0.18^{\circ}$\\
$LENGTH$&---&$0.18^{\circ}-0.38^{\circ}$\\
$CONCENTRATION$&$ < 0.25$&---\\
$D_{dist}$&$0.02^{\circ}-0.09^{\circ}$&$ < 0.12^{\circ}$\\
\hline
\hline
\end{tabular}
\end{center}
\end{table}

\begin{figure}
\centerline{\psfig{file=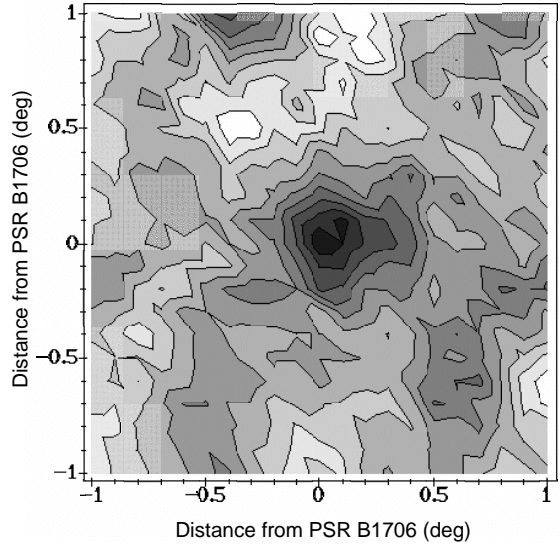,height=8cm}}

\caption{The results of a false source analysis for PSR B1706-44. The
grey scale is such that black corresponds to a detection probabilty of
$> 6~\sigma$. Contours are at $0.6~\sigma$ intervals}\label{lego}
\end{figure}

To demonstrate that this excess of $\gamma$-ray like events originates
from the source direction we have calculated the number of events
passing the {\em ALPHA\/}-cut for a number of different assumed source
positions (a `false source' analysis). We show the results of this
analysis in Figure~\ref{lego}. As the telescope is alt-azimuth mounted,
the field of view rotates around its centre with time; the effect of
this rotation has been allowed for in the analysis using a software
correction. Note that during these observations, the position of the
source in the telescope's field of view, known to an accuracy of
$0.02^{\circ}$, was displaced from the centre of the camera by
$\sim0.2^{\circ}$ degrees; this reduces the chance that the observed
excess is a result of a geometrical bias towards the camera centre. We
have also investigated the effects of the bright star $\eta$ Scorpii on
the alpha distribution, and conclude there are none \cite{Hold97}.

Fluxes have been estimated on the basis of the number of excess events
recorded with {\em ALPHA\/} $< 22.5^{\circ}$ divided by the time taken
to make the observations, which gives a $\gamma$-ray detection rate of
$0.26 \pm 0.05$ per minute. The $\gamma$-ray retention factor is
approximately 20 \%, giving an estimated incident $\gamma$-ray rate $
1.3 \pm 0.3$ per minute. The energy threshold for this object, which is
at an average zenith angle of $\sim~30^{\circ}$ is estimated to be 300
GeV. The collecting area of the telescope has been estimated using Monte
Carlo simulations and is $\sim 5.5 \times 10^{8}~{\rm cm}^{2}$ for these
observations. The integral flux from PSR B1706-44 is therefore $(3.9\pm
0.7_{\rm stat}) \times 10^{-11}~{\rm cm}^{-2}~{\rm s}^{-1}$ at $E > 300$
GeV, subject to a systematic error of $\sim 50 \%$. This corresponds to
a luminosity of $1.4 \times 10^{34}~{\rm erg~s}^{-1}$ at $E > 300$ GeV,
assuming a distance to the source of 2.8 kpc \cite{Kori95}, a spectral
index of $-2.8$, and that the emission is isotropic. This is $\sim
0.3\%$ of the total spindown energy of PSR B1706-44.

\subsection{Periodicity}

A pulsar timing analysis has been performed. The event times were
reduced to the Solar System Barycentre and analysed for periodicity
using the Rayleigh Test at the contemporary pulsar period derived from
the pulsar timing database maintained at Princeton
University\footnote{Available at web address
http://pulsar.princeton.edu/ftp/gro.}. We have no evidence for
periodicity from PSR B1706-44 in the dataset reported here. A $3~\sigma$
flux limit was calculated by equating the Rayleigh statistic
($\exp(-NR^{2})$ where $N$ is the number of events and $R$ is the fractional
strength of the pulsed signal) to the probability required for a
$3~\sigma$ detection. The 3 sigma flux limit for pulsed emission thus
calulated at less than 35\% of the observed $\gamma$-ray flux, which
corresponds to a flux limit of $\sim 1.4 \times 10^{-11}~{\rm
cm}^{-2}~{\rm s}^{-1}$.

\section{Discussion}

Our observations of VHE $\gamma$-ray emission from PSR B1706-44 confirm
the earlier reports by Kifune et al. \cite{Kifu95} that this
object emits in the VHE $\gamma$-ray waveband. The lowest energy events
recorded were $\sim 125$ GeV.

\begin{figure}
\centerline{\psfig{file=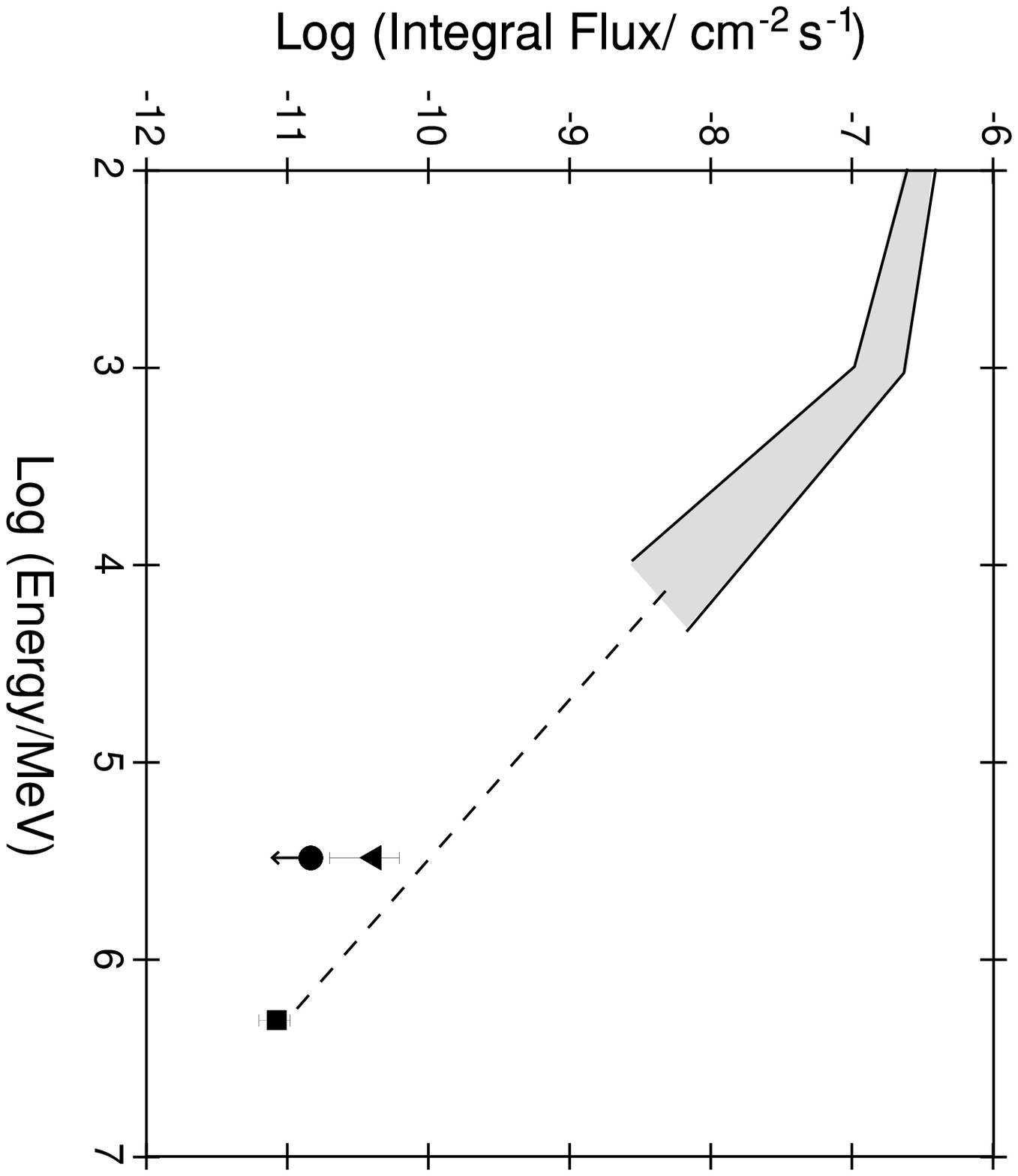,height=8cm,angle=90}}

\caption{The integral $\gamma$-ray spectrum of PSR B1706-44. The shaded
area indicates the EGRET spectrum for this object. $\blacktriangledown$
represents the unpulsed flux detected with the Mark 6 telescope, and
$\bullet$ represents the upper limit to the pulsed flux. $\blacksquare$
represents the measurement of Kifune et al. \cite{Kifu95}. Note that the
point of Kifune et al. has been moved to correspond with the revised
energy threshold of 2 TeV for CANGAROO at the time of the PSR B1706-44
measurements \cite{Robe97}.}\label{spectrum}

\end{figure}

Our dataset does not show evidence for pulsations at the known pulsar
frequency. This is in contrast to the result obtained with the EGRET
experiment, which suggest that $< 20\%$ of the total emission above 100
MeV can be ascribed to unpulsed emission, and that this fraction
decreases as energy increases \cite{Thom96}. Kifune et al. also find no
evidence for pulsed TeV emission. The flux measured with the Durham Mark
6 telescope is compatible with the extrapolation of the EGRET spectrum
above 1 GeV (see Figure~\ref{spectrum}). These measurements suggest that
the emission mechanisms at $1 {\rm~GeV} <~E < 5~{\rm GeV}$ and $E > 300$
GeV are very similar, and so it is significant that the proportion of
the $\gamma$-ray emission which is pulsed changes from greater than 80\%
to less than 35\%. It is of interest to investigate the lowest energy
events detected by the Mark 6 telescope as this opens up the possibility
of establishing emission at energies down to $\sim 125$ GeV and hence
identifying the range of energy within which the transition from pulsed to
unpulsed $\gamma$-radiation occurs. More data will be needed to increase
sensitivity at these low energies.

\begin{ack} We are grateful to the UK Particle Physics and Astronomy
Research Council for support of the project and the University of Sydney
for the lease of the Narrabri site. The Mark 6 telescope was designed
and constructed with the assistance of the staff of the Physics
Department, University of Durham. The efforts of Mr. P. Cottle, Mrs.
S.E. Hilton and Mr. K. Tindale are acknowledged with gratitude.
\end{ack}

\end{document}